**Langmuir and Langmuir-Blodgett films of a maleic anhydride derivative: effect of subphase divalent cations**

B. Martín-García, M. Mercedes Velázquez*

*Departamento de Química Física, Facultad de Ciencias Químicas. Universidad de Salamanca, E-37008-Salamanca, Spain.*

J. A. Pérez-Hernández

*Centro de Láseres Pulsados Ultraintensos (CLPU). E-37008-Salamanca, Spain*

and

J. Hernández-Toro

*Servicio Láser. Universidad de Salamanca, E-37008-Salamanca, Spain*

| | |
|---|---|
| Corresponding author: | M. Mercedes Velázquez |
| Corresponding address: | Departamento de Química Física |
| | Facultad de Ciencias Químicas |
| | Universidad de Salamanca |
| | Plaza de los Caídos s/n |
| | 37008 Salamanca, SPAIN |
| Fax: | 00-34-923-294574 |
| E-mail: | mvsal@usal.es |

**Abstract**

We report the study of the equilibrium and dynamic properties of Langmuir monolayers of poly (styrene-co-maleic anhydride) partial 2-buthoxy ethyl ester cumene terminated polymer and the effect of the $Mg(NO_3)_2$ addition in the water subphase on the film properties. Results show that the polymer monolayer becomes more expanded when the electrolyte concentration in the subphase increases. Dense polymer films aggregate at the interface. The aggregates are transferred onto Silicon wafers using the Langmuir-Blodgett methodology and the morphology is observed by AFM. The structure of aggregates depends on the subphase composition of the Langmuir film transferred onto the silicon wafer.

**Title running head**: Divalent cations on polymer Langmuir and Langmuir Blodgett films

**Introduction**

Polymers adsorbed to interfaces play an important role for basic materials science as well as for the design and development of technological applications such as adhesives, optical or protective coatings, and biosensors [1, 2]. So polymers are in large quantities applied as thin films or coatings where the stability, appearance or durability substantially depends on the surface properties. Monolayers of water insoluble polymers, Langmuir monolayers (LM), can be used to build ultrathin solid films by means of Langmuir-Blodgett (LB) methodology. These extremely thin films with high degree of structural order are used in a wide range of technologies such as electronic devices, UV and electron beam resists, or biosensors [3]. In addition, polymer films with thickness of nanometers provide ideal sample geometry for studying the effects of one dimensional confinement on the structure and dynamics of polymer molecules.

To construct LB films efficiently is necessary to know the equilibrium and rheological properties of Langmuir monolayers used for LB preparation. It is well established that the rheological properties are linked to the monolayer properties and play an important role in the transfer process of polymers from the air-liquid interface to solid supports [4]. Studies carries out with synthetic and natural polymers adsorbed at the interfaces showed that small changes on temperature [5,6,7] on polymer concentration [7-10] or structure [11,12] lead to significant changes on Langmuir monolayers and consequently, on the LB films generated from deposition of ordered monolayers from the air-water interface onto solid substrates [13].

On the other hand, the interactions between insoluble monolayers of surfactants and multivalent ions dissolved in the subphases have received a great interest because are of major importance in the manufacture of high-quality Langmuir-Blodgett films with potential applications in thin film technology [14]. When ions are present in the subphase, the surfactant monolayer usually becomes more ordered [15] and transfers more easily to a solid substrate (LB films) [16]. The presence of some divalent ions in the subphase results in the appearance of structures on monolayers [17]. These self-assembly systems are interesting from both, the fundamental point of view [18] and the technological applications [19, 20]. For many applications, nanomaterials must be embedded in a polymer matrix. These matrices provide chemical and mechanical stability and prevent the aggregation of nanomaterials [21,22]. In these cases polymers used as patterns often self-assembly at nanometer scale providing well-organized structures which play an important role in practical applications. The self-assembly structures have to be transferred onto solids to build nanodevices by using different techniques such as spin-coating, layer-by-layer assembly or Langmuir-Blodgett methodology. The Langmuir-Blodgett (LB) technique offers the possibility of preparing homogeneous thin films with well-defined layered structures and a very precise control of the thickness and composition [23]. On the other hand, small changes on composition, temperature, addition of salt in the subphases of the Langmuir films which serve as precursor of the LB films can improve the properties of the solid-supported polymer films. Therefore, to construct LB films efficiently characterization of the Langmuir films is required.

In order to gain insight the effect of Langmuir monolayer properties on the structure of polymer LB films, we have studied the effect of both, the polymer concentration and the addition of electrolyte in the subphase, on the properties of Langmuir and Langmuir-Blodgett films of a maleic anhydride polymer. The polymer chosen was poly (styrene-co-maleic anhydride) partial 2 buthoxy ethyl ester cumene terminated. We have chosen this polymer because some styrene/anhydride polymer derivatives have shown potential as electron beam resists [24]; consequently, it could be used as pattern for the fabrication of layered molecular electronic devices. Finally, to study the effect of the addition of salts on both, the LM and LB films, we have chosen a salt of divalent cation, magnesium nitrate, because divalent cations interact stronger than the monovalent ones with acid groups [14].

**Experimental section**

***Materials.*** The polymer poly (styrene-co-maleic anhydride) partial 2 buthoxy ethyl ester cumene terminated, PS-MA-BEE hereafter, was from Sigma-Aldrich, chart 1.

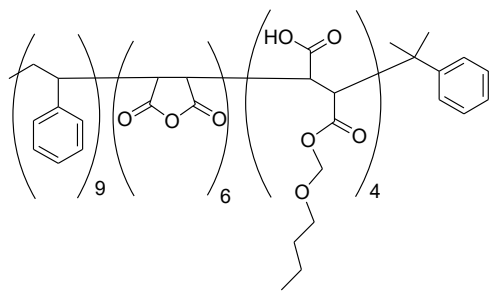

**Chart 1.** Molecular structure of poly (styrene-co-maleic anhydride) partial 2 buthoxy ethyl ester cumene terminated.

The ester: acid ratio provided by the manufacturer was 1:1. The polymer molecular weight $M_r$ = 2.5 kDa was also provided by the manufacturer. The polymer was used as received without further purification. The LB substrate is Si (100) wafer supplied by Siltronix (France). The silicon wafers were used without pretreatment.

Magnesium nitrate hexahydrate, 99% (A.C.S. reagent) and Chloroform (PAI, filtered) used to prepare the aqueous subphase and the spreading solutions, respectively, were from Sigma-Aldrich. Ultra pure water prepared using a combination of RiOs and Milli-Q systems from Millipore was used as subphase.

***Langmuir and Langmuir-Blodgett experiments.*** The pressure-area isotherms were recorded with a Langmuir mini-trough (KSV, Finland) placed on an anti-vibration table. Spreading solution of polymer in chloroform has a concentration of 0.06 mg ml$^{-1}$ and was prepared by weight using an analytical balance precise to ± 0.01 mg. Spreading solution was deposited onto the water subphase with a Hamilton microsyringe. The syringe precision is 1 µL. In order to confirm the film stability, two different modes for preparing the monolayers were used. In the first one the surface concentration was changed by subsequent additions of the polymer solutions on the liquid/air interface. The surface pressure was continuously monitored, and the equilibrium value was taken when the surface pressure Π had remained constant at least 10 min. In the second way, the monolayers were symmetrically compressed by moving two barriers under computer control after the spreading of the polymer solution. The maximum barrier speed was 2 mm/min. The surface pressure was measured with a Pt-Wilhelmy plate connected to an electrobalance. The temperature of the different subphases was maintained by flowing thermostated water through jackets at the bottom of the trough. The temperature was controlled by means of a thermostat/cryostat Lauda Ecoline RE-106. The temperature near the surface was measured with a calibrated sensor from KSV. A KSV2000 System 2 for Langmuir-Blodgett deposition was also used. The monolayer was transferred to solid substrates by compression at a barriers speed of 5 mm/min, with the substrate into the trough by vertically dipping it up at 5 mm/min.

The relaxation experiments were performed on the Langmuir mini-trough by the step-compression experiments [5]. The barriers were moved after the desired surface concentration has been reached and the equilibrium value of π obtained. In these experiments the area change, ΔA, remained in the linear regime. The time necessary to make the compression was slightly less than 2 s. Reading of π at constant intervals of 1s was taken until the equilibrium π-value corresponding to the final area was reached. The relaxation curve is accepted only if the initial and final values of surface pressure for a given compression agree with those corresponding to equilibrium.

***Surface potential measurements***. The surface potential ΔV was measured in the Teflon Langmuir minitrough with a Kelvin probe SPOT1 from KSV (KSV, Finland) located at approximately 2 mm above the aqueous surface. The Kelvin probe is based on the non-contact vibrating plate capacitor method with the reference electrode placed in the subphase. The surface potential of monolayers was determined relative to the surface potential of the supporting electrolyte and of water if no salt was added to the subphase. Each reported value is an average over five measurements and the standard deviation of these measurements was considered the experimental error.

***Atomic Force Microscopy (AFM).*** AFM images of the Langmuir-Blodgett (LB) film on silicon substrates were obtained in constant repulsive force mode by AFM (Nanotec Dulcinea, Spain) with a rectangular microfabricated silicon nitride cantilever (Olympus OMCL-RC800PSA) with a height of 100 μm, a Si pyramidal tip and a spring constant of 0.73 N/m. The scanning frequencies were usually in the range of 0.5 and 2.0 Hz per line. The measurements were carried out under ambient laboratory conditions. The images were obtained with the WSXM 5.0 program [25].

*Ellipsometry.* Multiangle measurements were performed on a null ellipsometer (Nanofilm, model EP3, Germany) to determine the refractive index and the thickness of the Langmuir-Blodgett LB films ~~better~~. The incident Nd-YAG laser beam ($\lambda$ = 532 nm) was focused on the LB films on silicon substrates positioned on the goniometer plate. The interface was modeled with the three layers model previously described [26]. The precision on the ellipsometric angles is ca. 0.0006.

**Results and discussion**

*Equilibrium properties of polymer monolayers.* Figure 1 shows the surface pressure vs. the surface concentration isotherms for PS-MA-BEE monolayers at 23ºC prepared in water and saline subphases. The isotherms were obtained by equilibration after each addition. For the sake of comparison isotherms obtained by continuous compression are represented in Figure 1 by lines.

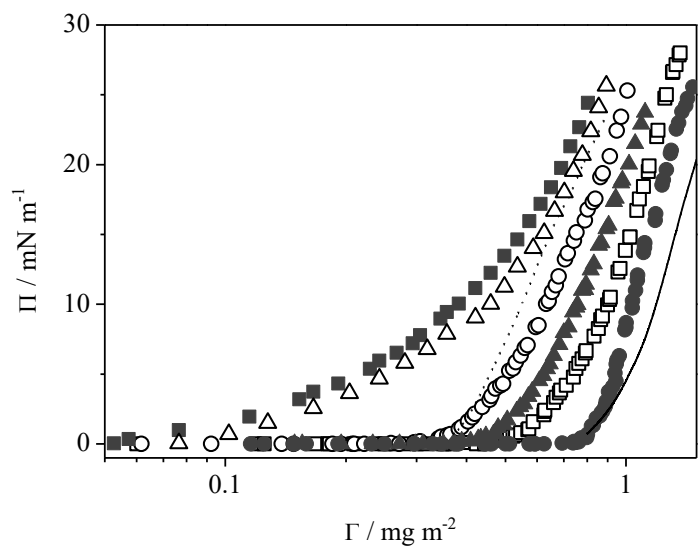

**Figure 1.** Surface pressure isotherms of PS-MA-BEE monolayers prepared in different subphases: (solid circles) water; (open squares) 0.12M Mg(NO$_3$)$_2$; (triangles) 0.18 M Mg(NO$_3$)$_2$; (open circles) 0.24M

Mg(NO$_3$)$_2$ ; (open triangles) 0.40 M Mg(NO$_3$); (solid squares) 0.50 M Mg(NO$_3$). Surface isotherms were obtained by addition (symbols) and by symmetric compression (dot and solid lines).

Results in Figure 1 show that monolayers in saline solutions are more expanded than monolayers in unsalted subphase. This effect increases as the electrolyte concentration in the subphase increases until it reaches the value of 0.4 M. The further addition of Mg(NO$_3$)$_2$ in the aqueous subphases does not modify the isotherm.

Figure 1 also shows that the isotherms obtained for addition and for continuous compression agree with each other until a given surface pressure. Above this pressure the monolayer obtained by compression is far from the equilibrium state and two opposite trends are observed as a function of the subphase composition; thus, in water subphase, the surface pressure values for monolayers prepared by compression are lower than the values corresponding to the monolayers obtained by addition, while for monolayers prepared in saline subphases, the surface pressure obtained by compression is higher than the addition one. In the later if the barriers stop, the surface pressure decreases until the value corresponding to the equilibrium. These facts can be due to the existence of dynamic processes on the monolayer [27-29]. In order to gain some insights into this dynamic process, the step-compression combined with oscillatory barriers experiments were carried out. These experiments can be carried out in a Langmuir surface balance and they allow obtain information of slow relaxation processes in monolayers. Results will be presented in the second part of this manuscript.

It is also necessary to check the stability of monolayers prepared by addition, with this purpose the following test was done: after addition of a given volume of the spreading solution, the surface pressure was monitored at least 4h. In all monolayers the surface pressure was found constant within the experimental uncertainty, indicating stable monolayers. Therefore, we choose the addition procedure to obtain stable polymer monolayers.

Figure 2 show the electric surface potential isotherms of polymer monolayers in water and in saline subphases.

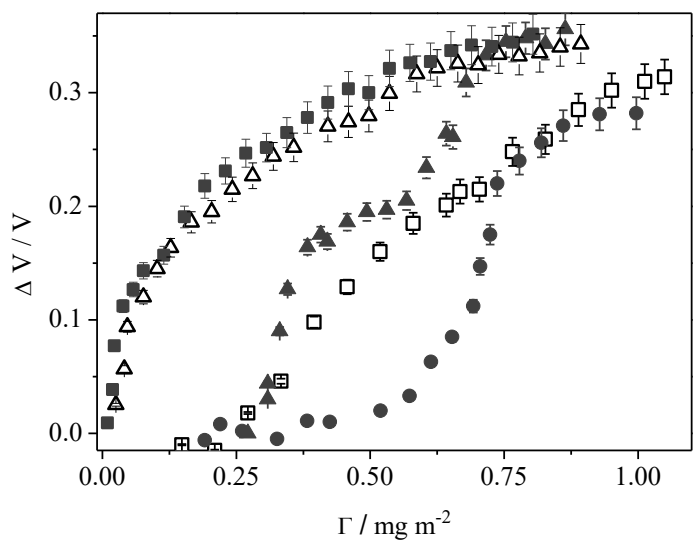

**Figure 2.** Surface potential isotherms at 296 K of PS-MA-BEE monolayers prepared in different subphases: (solid circles) water; (open squares) 0.12M $Mg(NO_3)_2$; (solid triangles) 0.24M $Mg(NO_3)_2$ (open triangles); 0.40M $Mg(NO_3)_2$ and (squares) 0.50M $Mg(NO_3)_2$

The surface potential values are always positive and for a given surface concentration they increase as the electrolyte concentration in the subphase increase. The effect is more marked in the dilute surface concentration regime.

As can be observed the surface potential curves have similar trend to the surface pressure ones; thus, the surface potential isotherms are shifted to lower surface concentration when the electrolyte concentration in the subphase increases. The isotherm in 0.24 M $Mg(NO_3)_2$ shows a pseudo-plateau while it is marked by a kink in the isotherms at lower electrolyte concentration in the subphase. The plateau almost disappears when the monolayer is built in saline subphases containing electrolyte concentrations above 0.4M.

It is also interesting to notice that the surface potential at high surface coverage does not increase with the electrolyte concentration for monolayers containing high electrolyte concentration and the value remains constant in a value of 0.35 V. This fact indicates that in these monolayers all acid groups are dissociated.

Figure 3 presents the equilibrium elasticity modulus for different polymer monolayers. The elasticity modulus is obtained from the pressure isotherms and the following equation $\varepsilon_0 = \Gamma \left( \dfrac{\delta \Pi}{\delta \Gamma} \right)$.

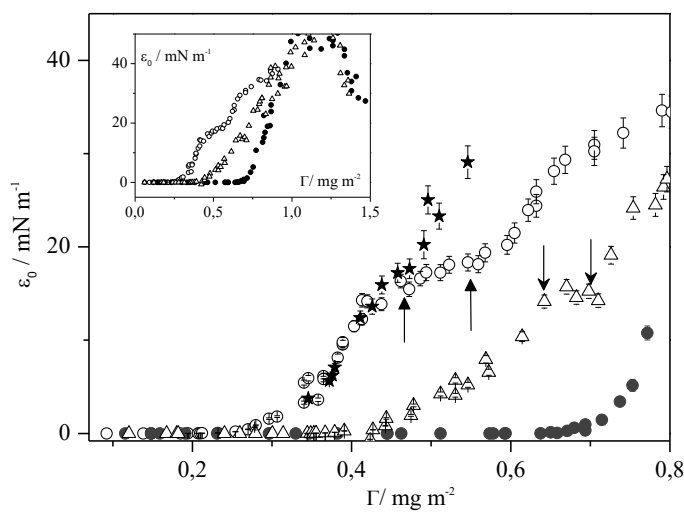

**Figure 3**. Variation of the equilibrium elasticity with polymer surface concentration for monolayers prepared in different subphases at 296K: (solid circles) water; (triangles) 0.12M Mg(NO$_3$)$_2$; (open circles) 0.24 M Mg(NO$_3$)$_2$ . The arrows point polymer concentrations at which the kink in the surface potential isotherms appears. Stars represent the dilational modulus obtained from step-compression experiments, see text.

Small elasticity values are observed for monolayers in water until the surface concentration reaches a value of 0.8 mg m$^{-2}$. A sudden increase of elasticity is observed above this surface concentration until it reaches a value of 50 mN m$^{-1}$. Finally, when the surface concentration is further increased a weak decrease of the elasticity is observed. This behavior is characteristic of films formed exclusively by polymer molecules, concentrated regime. In these concentrated films the conformational degrees of freedom of the polymer coils decrease and consequently, the elasticity modulus also decreases [5, 29]. It is also interesting to notice that the equilibrium elasticity modulus values are consistent with liquid-expanded surface states [30] in all these monolayers. On the other hand, and because the isotherms obtained by compression do not correspond to the equilibrium ones (isotherms prepared by addition), the elasticity values calculated from the isotherm obtained by compression are far from the equilibrium elasticity values, see figure 3.

Differences between the elasticity modulus of monolayers in saline and in the unsalted subphases are the following: a) the equilibrium elasticity values for dilute monolayers ($\Gamma < 1$ mg m$^{-2}$) are higher for monolayers in saline subphase than in the subphase without salt. The effect is more marked when the salt concentration increases. b) The equilibrium elasticity curves for monolayers in saline subphases with salt concentrations below 0.4M, present a kink or a plateau at the same surface concentration values than in the surface potential curves. For comparative purposes the arrows in Figure 3 point to the position of the pseudo-plateau in the surface potential isotherms.

The surface potential and pressure isotherms of PS-MA-BEE clearly show that the monolayers become more expanded when the Mg(NO$_3$)$_2$ concentration in the subphase increases until it reaches a value of 0.4M. Above this concentration the isotherm remains unaffected by the increase of electrolyte concentration. This behavior could be attributed to the formation of a complex between the divalent cation $Mg^{2+}$ and the acid groups of the polymer molecules. The isotherm expansion was observed by other authors when metal-surfactant complexes are formed at the interface and the surfactant contains bulky head groups [31]. This could be the scenario in our system because the hydrophilic part of the polymer is too bulky and some molecular rearrangements have to be carried out in order to form the complexes between the divalent cation and the carboxylic groups of the polymer. Results also show that the isotherm expansion stop when the $Mg(NO_3)_2$ concentration in the subphase reaches a value of 0.4M. This fact indicates that at this salt concentration all the polymer molecules are bound to the divalent cation.

The coincidence between the kink and the plateau positions observed in the surface potential and the equilibrium elasticity isotherms may be indicative of phase transitions that in this system can be originate by different populations of free and bound polymer molecules. According to it, the plateau disappears when the cation-polymer complexes at the interface predominate, $[Mg(NO_3)_2] \geq 0.4M$.

***Dynamic properties of polymer monolayers***. We study the relaxation process of the monolayer using the step-compression method [5]. In this method, the monolayer is perturbed by a quick compression, and then the relaxation to the new equilibrium condition is followed by surface pressure measurements. Previously, we also use oscillatory barrier measurements to check the strain range where the amplitude of the stress, $\sigma_0$, is proportional to the strain, $U_0 = \Delta A/A$, the linear regime. Examples of oscillatory experiments are presented in Figure 4 and are preformed at a frequency of 0.006 Hz. For the sake of clarity Figure 4 shows the results obtained for two different subphases, water and 0.24M $Mg(NO_3)_2$.

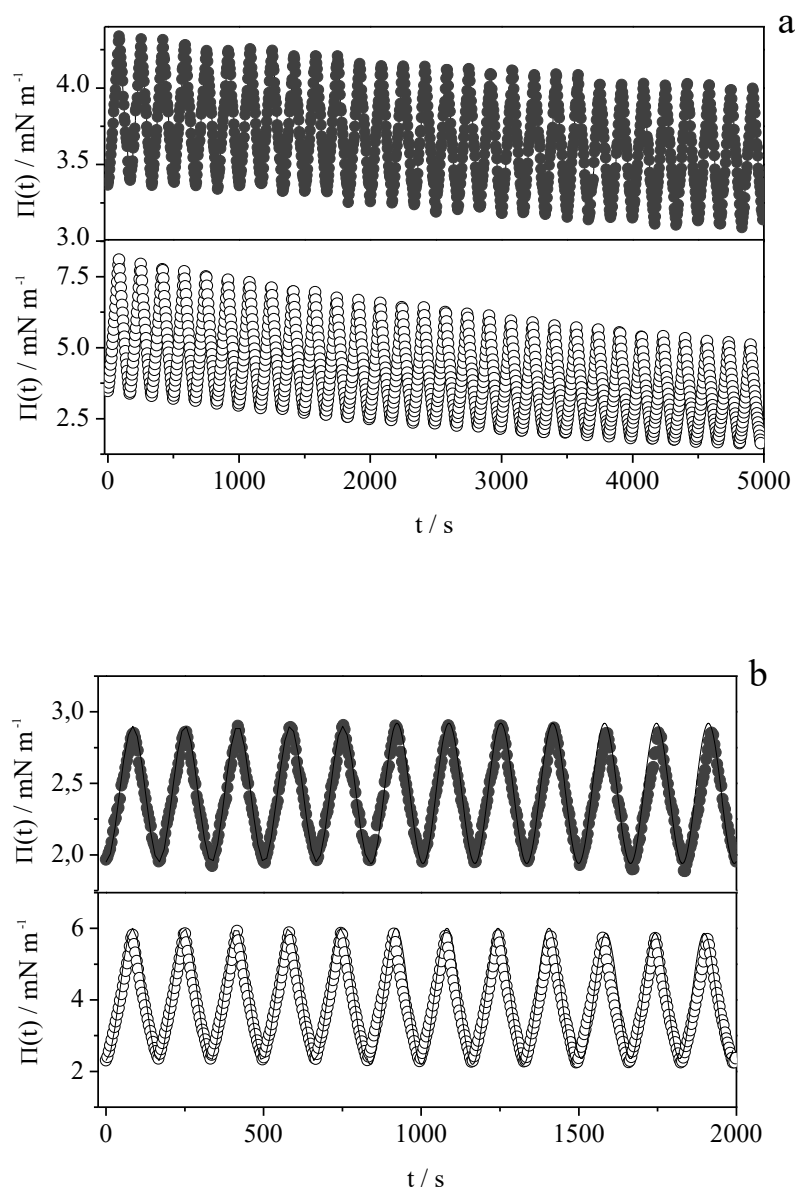

**Figure 4.** Oscillatory experiment results at 296 K for different monolayers: (a) pure water and polymer surface concentration around 0.9 mg m$^{-2}$ ; b) 0.24M $Mg(NO_3)_2$, and polymer surface concentration around 0.42 mg m$^{-2}$. The strain amplitude was 4.3 % (solid circles) and 12 % (open circles). The experiments were carried out at the frequency of 0.006 Hz.

As can be seen in Figure 4a for monolayers in water, after each compression-expansion cycle, the former initial state is not recovered and the pressure decreases after each cycle. The amplitude of the stress remains almost constant at low strains while it decreases at high strains. This behavior is not observed in monolayers prepared in saline subphases and can be explained by developing of irreversible features in the film as it is strained by subsequent compression-expansion cycles. This behavior has been observed in other polymer monolayers [32]; however, this fact could be simply due to the desorption process followed by the polymer dissolution in the subphase. To disregard this possibility we carried out the following experiment: a polymer monolayer prepared by addition was submitted to compression-expansion cycles, then the barriers were stopped and the pressure monitored with time (2h). The surface pressure decrease was least of 1%. Similar results were obtained at different surface coverage. This observation is compatible with no polymer dissolution in the subphase when the monolayer is strained.

We are interested to study the relaxation process of polymer monomers adsorbed in the monolayer; therefore we have only studied the dynamic processes on the monolayers prepared on saline subphases. The inset in Figure 5 presents the variation of the amplitude of the stress $\sigma_0$ with the strain, $U_0$ for monolayers in saline subphases. As can be seen in the figure, deviations of the Hookean linear behavior are observed for the area change above 10% of the total area. Consequently, in the step-compression experiments the area change, $\Delta A$, is kept below 10% in order to ensure that the system remains in the linear regime. This nonlinear behavior has been found for other polymer monolayers [6, 12].

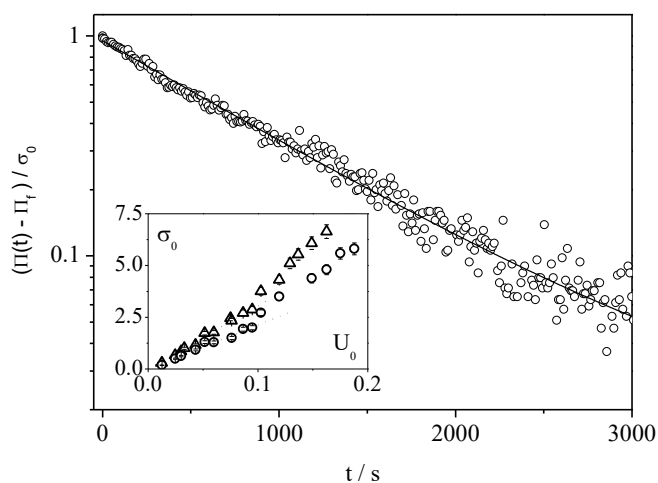

**Figure 5.** Stress-relaxation experiment for a polymer monolayer in 0.24M $Mg(NO_3)_2$ subphase with a surface concentration of 0.38 mg m$^{-2}$ at 296 K. Line is the single exponential curve. The inset shows the variation of the amplitude of the stress, $\sigma_0$, with the strain, $U_0$ for polymer monolayers prepared in 0.12M $Mg(NO_3)_2$ subphase (triangles) and in 0.24M $Mg(NO_3)_2$ (circles). The dot lines represent the Hookean's linear behavior.

Stress-relaxation experiments show that in the low-coverage region, before the end of the kink or the pseudo-plateau in the isotherms, the stress-relaxation curve can be described by a single exponential function, Figure 5. However, when the surface coverage is further increased no single-exponential relaxations are found, this fact is indicative of additional processes in the monolayer. In monolayers prepared in the most concentrated saline subphases, $Mg(NO_3)_2 \geq 0.4$ M, the stress-relaxation curves are single-exponential functions for surface concentrations below 1 mg m$^{-2}$. Above this surface concentration no single-exponential relaxations were found.

Figure 6 shows the concentration dependence of the relaxation time obtained for different monolayers. For the sake of clarity only some results are presented in Figure 6. As can be seen in the figure different trends are found. For monolayers in saline subphases with salt concentrations below 0.4M, at low-coverage, before the kink or the pseudo-plateau, the relaxation time increases with the polymer concentration, this behavior is characteristic of slow collective motions of the polymer coils adsorbed at the interface [31, 32]. Conversely, at high-coverage the relaxation time decreases when the polymer surface concentration increases. According to the equilibrium properties, these surface concentrations correspond to regions in the isotherm in the phase transitions; consequently, the concentration dependence of the relaxation time is a further indication of a possible phase transition between a state in which we have polymer molecules bound to the $Mg^{2+}$ and a state with free polymer molecules.

On the other hand, when monolayers are prepared in subphases containing electrolyte concentrations above 0.4 M, the stress-relaxation experiments have led to relaxation curves described by a single exponential in the surface concentration range between 0.15 and 1 mg m$^{-2}$ and the relaxation time increases with the surface concentration. As can be seen in Figure 6, the relaxation time increases as the surface coverage and the values found are quite similar to those corresponding ones obtained for monolayers in 0.24 M of $Mg(NO_3)_2$ and at low surface coverage.

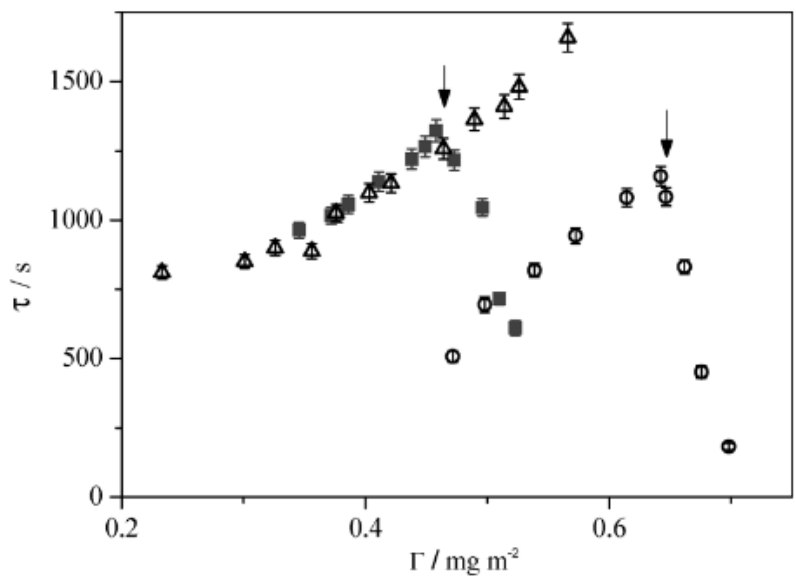

**Figure 6.** Variation of the relaxation time with the polymer surface concentration for monolayers prepared in the following subphases at 296K: (open circles) 0.12M $Mg(NO_3)_2$; (solid circles) 0.24M $Mg(NO_3)_2$; and (solid squares) 0.50 M $Mg(NO_3)_2$. The arrows point polymer concentrations at which the kinks in the surface potential and equilibrium elasticity isotherms appear.

We have also determined the dilatational modulus, $|\varepsilon|$, from the relaxation curves as: $|\varepsilon| = \dfrac{\sigma_0}{U_0}$. Some of these values are represented in Figure 3 by stars. For the sake of clarity we only represent the values for monolayers in subphases with 0.24 M and in 0.5 M of $Mg(NO_3)_2$. Results in Figure 3 show that in the monolayers prepared in the subphase containing $Mg(NO_3)_2$ 0.24 M, there is an excellent agreement between the dilatational modulus and the equilibrium elasticity at low-coverage. However, when the coverage increases the dilatational modulus is higher than the equilibrium elasticity, this is indicative of viscoelastic behavior. The crossover between the elastic and viscoelastic behavior appears at the polymer surface concentration where deviations between the isotherms obtained by addition and compressions become important. This behavior is observed for all

monolayers with electrolyte concentrations below 0.4M; however, when the electrolyte concentration in the subphase increases, $[Mg(NO_3)_2] \geq 0.4M$, the dilatational modulus values are always higher than the equilibrium elasticity ones. The differences between dilatational and equilibrium elasticity values are quite marked. This fact indicates that the formation of complexes between the polymer and the divalent cation increases the viscoelastic behavior of the monolayer.

As it was indicated above the stress-relaxation experiments in the densest monolayers have led to relaxation curves described by no single-exponential curves that can be signature of additional processes such as disorder-order transitions [12] or polymer aggregation. To disregard this point we transfer polymer monolayers from the liquid interface to Si (100) via the Langmuir-Blodgett technique (LB) because several studies shown that aggregates produced by self-assembly monolayers can survive when are transferred from the Langmuir monolayer to solid supports [33-35]. For comparative purposes we have also constructed LB films from Langmuir monolayers with low surface coverage, below the end of the kink or of the pseudo plateau. The AFM images shown no domains in these LB films (not shown).

According to the differences observed in the properties of Langmuir films prepared by addition and by compression, we build the LB films by transferring Langmuir films prepared by the two different methodologies. Different mesoscopic structures are detected as a function of the methodology used. To illustrate this fact several AFM images are collected in Figure 7. We have checked the reproducibility of the transfer procedure by obtaining each LB film at least three times. The results agree with each other except for LB films built by transferring the Langmuir film prepared by addition in a water subphase.

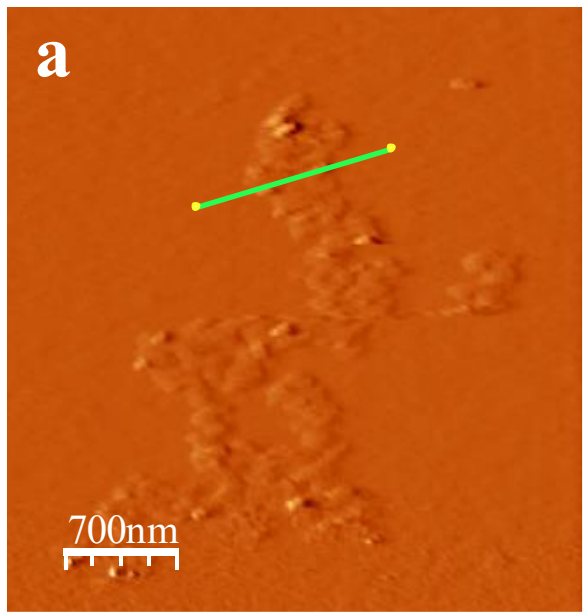
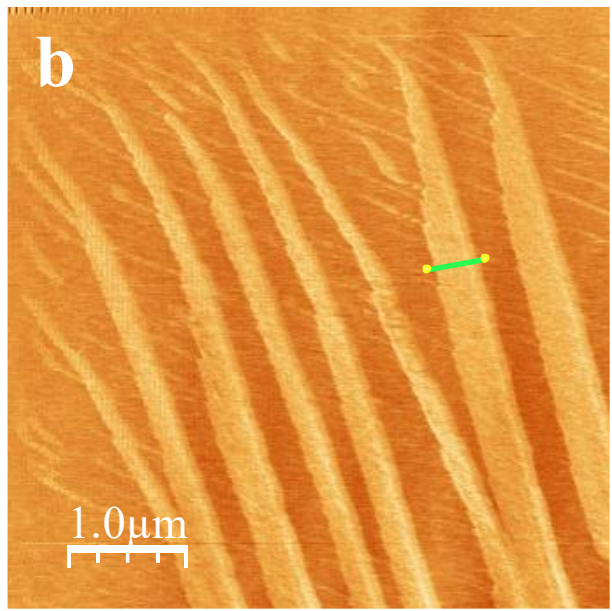
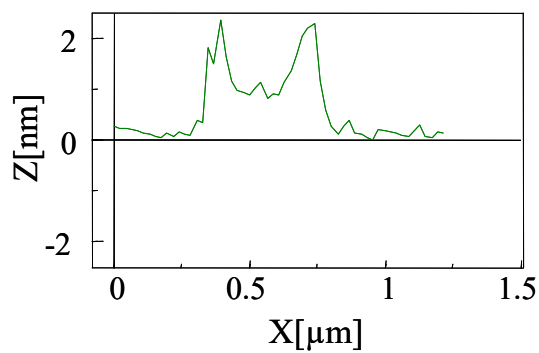
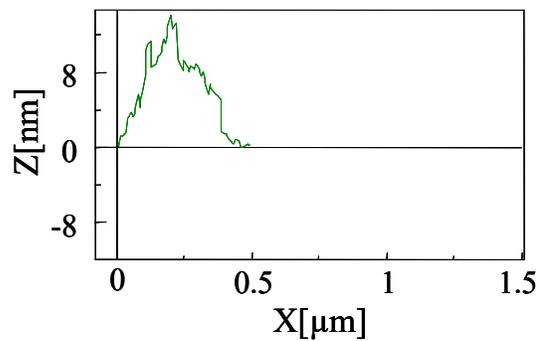
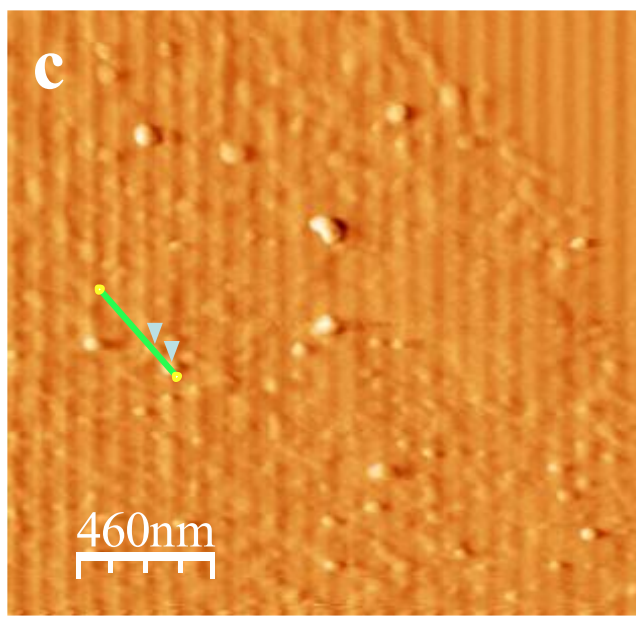
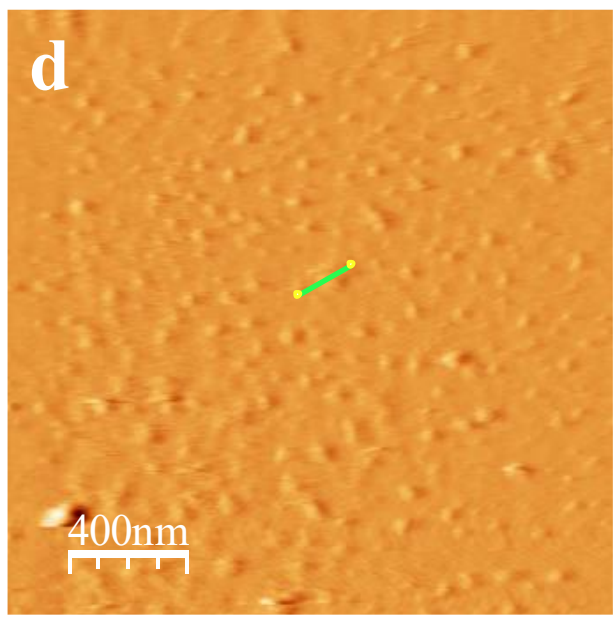
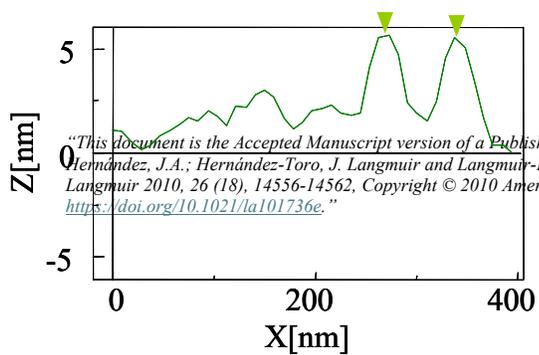
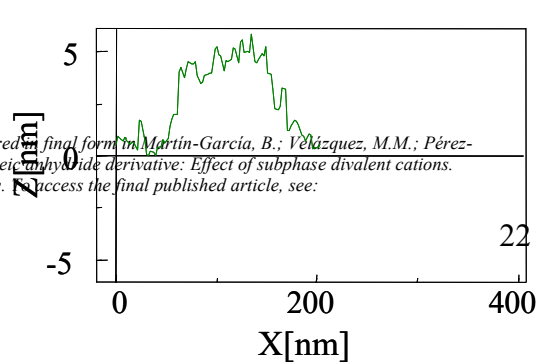

**Figure 7.** AFM images of the LB films on Si (100) for different PS-MA-BEE monolayers: in water subphase at surface pressure 14 mN/m and 0.12M Mg(NO$_3$)$_2$ at 10 mN/m, by subsequent additions (a, c) and by compression (b, d), respectively. Below are the corresponding cross section profiles. In addition the green angle pairs correspond to the angles pairs in the figure 7c.

Images in Figure 7 clearly show different structures as a function of both, the subphase composition and the methodology employed to prepare the Langmuir monolayer which serve to build the LB films. Thus, when the Langmuir monolayer was prepared in aqueous subphase by compression, long stripes are observed, Figure 7b. These aggregates present a high degree of reproducibility. In contrast, aggregates with different morphologies are observed when the Langmuir monolayer is obtained by addition in water subphase. In this case the morphology of these aggregates also changes when the polymer concentration of the spreading solution changes. This behavior is consistent with results obtained in previous works for diblock copolymers with immiscible blocks confined to the surface [37]. A theoretical model has been proposed to interpret this fact [28]. The model considers that the fast evaporation of the spreading solvent and the extremely low solubility of the hydrophobic blocks compared with the hydrophilic ones are the main reasons of the association in metastable aggregates. This behavior depends on the concentration of the spreading solution.

The figures 7c and 7d, show images of LB films built from dense Langmuir monolayers in saline subphases 0.12M of Mg(NO$_3$)$_2$ prepared by addition and by compression methodologies, respectively. Even though the morphology of aggregates acceptably agrees with each other, differences in the monodispersity degree can be observed; thus, LB's built by transferring Langmuir films prepared by compression are almost monodisperse and the average value of the roughness is around 5 nm. In contrast, the LB prepared by addition show higher polidispersity degree and the roughness values are between 3 and 5 nm. Finally, it is interesting to notice that results in saline subphases obtained by the two methodologies are reproducible.

Although the images in figure 7 correspond to Langmuir-Blodgett built with Langmuir films in 0.12M $Mg(NO_3)_2$ subphase; similar images are observed for the rest of films prepared in saline subphases. From the AFM results it is possible to conclude that aggregates transferred from the concentrated monolayers of polymer have different morphology than the ones transferred from monolayers containing polymer-cation complexes.

Finally, we have also used a laser ellipsometer to obtain the layer thickness of the different LB films. In order to obtain the film thickness, different approaches have been used to model the interfases. We have modeled these films as three layers: the first one is the Si substrate with a refractive index of 4.1264; the second layer is the native oxide layer of a refractive index of 1.4653 [37]. The thickness of this layer was obtained in an ellipsometry experiment using the clean silicon wafer. Finally, the third layer (polymer film) thickness and the refractive index are obtained from the fitting of the two ellipsometric angles, $\Psi$, and $\Delta$, to Fresnel equation. Results obtained are collected in Table 1. From data in Table 1 it is possible to conclude that the thickness values obtained by ellypsometry are in good agreement with the average roughness of films obtained by AFM measurements.

**Table 1.** Thickness and roughness values obtained from Ellipsometry and AFM measurements, respectively.

| Subphase | π /mN m$^{-1}$ | LB Method | Roughness AFM (nm) | Thickness Ellipsometry (nm) |
|---|---|---|---|---|
| Water | 14 | Addition | 1.5 – 2.5 | 2.5 ± 0.7 |
|  | 14 | Compression | 8.0 ± 0.6 | 8.1 ± 2.4 |
| Mg(NO$_3$)$_2$ 0.12 M | 3.8 | Addition | 1.0 ± 0.5 | 1.8 ± 0.5 |
|  | 10 | Addition | 3-5 | 2.9 ± 0.8 |
|  | 10 | Compression | 5.0 ± 0.5 | 4.4 ± 1.3 |

**Conclusions**

We were interested to study the effect of electrolytes in the subphase on the equilibrium and dynamic properties of Langmuir and Langmuir-Blodgett films of the polymer poly (styrene-co-maleic anhydride) partial 2 buthoxy ethyl ester cumene terminated, in order to prepare films supported with potential applications in the construction of electronic devices. We choose this polymer because has showed potential as electron beam resists and, consequently, it could be used as pattern in the preparation of nanodevices. Results in this work show that the Langmuir films in saline solutions are more expanded and more stable than the monolayer without salt in the subphase. These facts can be attributed to the formation of complexes between the divalent cation and the carboxylic groups of the polymer molecule. Results found in this work also demonstrate self-assembly at the interface at high surface coverage. The aggregates can be transferred from the air-liquid interface onto silicon wafers. Different morphologies are observed by AFM as a function of both, the method to build the LB film and the subphase composition. The thickness of LB films obtained by ellypsometry agrees very well with the roughness determined by the AFM measurements.

**Acknowledgments**

The authors thank financial support from ERDF and MEC (MAT 2007-62666) and from Junta de Castilla y León (SA138A08). B.M.G. wishes to thank Junta de Castilla y León for her FPI grant. We are grateful to the UIRC of the Universidad Complutense de Madrid for making available the ellipsometry facility, in particular, J.E.F. Rubio for the ellipsometric measurements. Drs. F. Ortega and R.G. Rubio of the Complex Systems Group, GSC, (Universidad Complutense de Madrid) are also acknowledged by helpful discussions of results. We also thank to Ultra-Intense Lasers Pulsed Center of Salamanca (CLPU) for the AFM measurements.

**References**

(1) Jones, R.A.L.; Richards, R.W. *Polymers at Surfaces and Interfaces*; Cambridge University Press: U.K., 1999.

(2) Stamm, M. (Ed.) *Polymer Surfaces and Interfaces: Characterization, Modification and Applications;* Springer-Verlag: Berlin, 2008.

(3) Petty, M.C. *Thin Solid Films* **1992**, *210/211*, 417-426.

(4) Regismond, S.T.A.; Winnik, F.M.; Goddard, E.D. *Colloid Surf. A: Physicochem. Eng. Aspects.* **1996**, *119*, 221-228.

(5) Hilles, H.M.; Ortega, F.; Rubio, R.G.; Monroy, F. *Phys. Rev. Lett*. **2004**, *92*, 255503-255507.

(6) Hilles, H.M.; Maestro, A.; Monroy, F.; Ortega, F.; Rubio, R.G.; Velarde, M.G. *J. Chem. Phys*. **2007**, *126*, 124904-1-10.

(7) Hilles, H.M.; Ritacco, H.; Monroy, F.; Ortega, F.; Rubio, R.G. *Langmuir* **2009**, *25*, 11528-11532.

(8) Monroy, F.; Hilles, H.M.; Ortega, F.; Rubio, R.G. *Phys. Rev. Lett*. **2003**, *91*, 268302-1.

(9) Hilles, H.M.; Sferrazza, M.; Monroy, F.; Ortega, F.; Rubio, R.G.; Velarde, M.G. *J. Chem. Phys*. **2006**, *125*, 074706-074707.

(10) Petkov, J.T.; Gurkov, T.D.; Campbell, B.E.; Borwankar, R.P. *Langmuir* **2000**, *16*, 3703-3711.

(11) Luinge, J.W.; Nijboer, G.W.; Hagting, J.G.; Vorenkamp, E.J.; Fuller, G.G.; Schoute, A.J. *Langmuir* **2004**, *20*, 11517-11522.

(12) Miranda, B.; Hilles, H.M.; Rubio, R.G.; Ritacco, H.; Radic, D.; Gargallo, L.; Sferrazza, M.; Ortega, F. *Langmuir* **2009**, *25*, 12561-12568.

(13) Chen, X.; Lenhert, S.; Hirtz, M.; Lu, N.; Fuchs, H.; Chi, L. *Acc. Chem. Res*. **2007**, *40*, 393-401.

(14) Roberts, G.G. *Langmuir-Blodgett films*; Plenum Press: New York, 1990.

(15) Schwartz, D.K. *Surf. Sci. Rep*. **1997**, *27*, 241-334.

(16) Ghaskadvi, R.S.; Carr, S.; Dennin, M. *J. Chem. Phys*. **1999**, *111*, 3675-3678.

(17) Mann, S. *Science* **1993**, *365*, 499-505.

(18) Bloch, J.M.; Yun, W. *Phys. Rev. A*. **1990**, *41*, 844-862.

(19) Knipping, E.M.; Lakin, M.J.; Foster, K.L.; Jungwirth, P.; Tobias, D.J.; Gerber, R.B.; Dabdub, D.; Finlayson-Pitts, B.J. *Science* **2000**, *288*, 301-306.

(20) Laskin, A.; Gaspar, D.J.; Wang, W.; Hunt, S.W.; Cowin, J.P.; Colson, S.D.; Finlayson-Pitts, B.J. *Science* **2003**, *301*, 340-344.

(21) Tomczak, N.; Janczweski, D.; Han, M.; Vancso, G.J. *Prog. Polymer Sci*. **2009**, *34*, 393-430.

(22) Kim, H-C.; Park, S-M.; Hinsberg, W.D. *Chem. Rev*. **2010**, *110*, 146-177.

(23) Petty, M.C. *Langmuir–Blodgett Films.: An Introduction;* Cambridge University Press: Cambridge, 1996.

(24) Jones, J.; Winter, C.S.; Tredgold, R-H. Hodge, P., Hoorfar, A. *Polymer* **1987**, *28*, 1619-1626.

(25) Horcas, I.; Fernández, R.; Gómez-Rodríguez, J.M.; Colchero, J.; Gómez-Herrero, J.; Baro, A.M. *Rev. Sci. Instrum*. **2007**, *78*, 013705-1-8.

(26) Tompkins, H.G. *A User's Guide to Ellipsometry*; Academic Press Inc.: London,

1993.

(27) Li, S.; Clarke, C.J.; Eisenberg, A.; Lennox, R.B. *Thin Solid Films* **1999**, *354*, 136-141.

(28) Hosoi, A.E.; Kogan, D.; Deveraux, C.E.; Bernoff, A.J.; Baker, S.M. *Phys. Rev. Lett*. **2005**, *95*, 037801-1-4.

(29) López-Díaz, D.; Velázquez, M.M. *Eur. Phys. J. E.* **2008**, *26*, 417-425.

(30) Davies, J.T.; Riedel, E.K., in *Interfacial Phenomena*; Academic Press: New York, 1963, p. 265.

(31) Jayathilake, H.D.; Driscoll, J.A.; Bordenyuk, A.N.; Wu, L.; da Rocha, S.P.R.; Verani, C.N.; Benderski, A.V. *Langmuir* **2009**, *25*, 6880-6886.

(32) Hilles, H.; Monroy, F.; Bonales, L.J.; Ortega, F.; Rubio, R.G. *Adv. Colloid Interface Sci.* **2006**, *122*, 67-77.

(33) Liao, Q.; Carrillo, J-M.Y.; Dobrynin, A.V.; Rubinstein, M. *Macromolecules* **2007**, *40*, 7671-7679.

(34) Burbi, M.K.; Malik, A.; Richter, A.G.; Huang, K.G.; Dutta, P. *Langmuir* **1997**, *13*, 6547-6549.

(35) Meine, K.; Vollhardt, D.; Weidenmann, G. *Langmuir* **1998**, *14*, 875-879.

(36) Hamley, I.W. *The Physics of Diblock Copolymers*; Oxford: New York, 1998.

(37) Palik, E.D. *Handbook of Optical Constant of Solid*; Academic Press Inc.: San Diego, 1993.